\documentclass[referee]{aa}

\usepackage{graphicx}
\usepackage{amssymb}
\usepackage{natbib}
\usepackage{epsf}

\renewcommand{\ion}[2]{#1\,{\sc #2}}

\newcommand{\lam}{$\lambda$}

\newcommand{\ecss} {erg~cm$^{-2}$~s$^{-1}$~sr$^{-1}$} % erg/cm2/s/sr
\newcommand{\deltE}{\Delta\kern-1ptE}

\newcommand{\asec}{$^{\prime\prime}$}

\title{The element abundance FIP effect in the quiet Sun}

\author{P.R.\ Young}

\institute{CCLRC Rutherford Appleton Laboratory, Chilton, Didcot,
  Oxfordshire, OX11 0QX, U.K.}

\date{Received / Accepted}

\abstract{
The Mg/Ne abundance ratio in the quiet Sun is measured in
both network and supergranule cell centre regions through EUV spectra
from the Coronal Diagnostic Spectrometer on SOHO. Twenty four sets of
data over 
the period 1996 March to 1998 June (corresponding to solar minimum)
are studied. Emission lines of the sequences Ne IV--VII and Mg
V--VIII are simultaneously analysed by comparing with theoretical
emissivities from the CHIANTI database to yield the Mg/Ne abundance
and emission measure over the temperature region $5.0\le\log\,T\le
6.1$. The average enhancements over the photospheric Mg/Ne abundance
are found to be $1.25\pm 0.10$ (network) and $1.66\pm
0.23$ (cell centres), significantly lower than the typical 4--5
enhancements  found in the slow solar wind. This result implies that
only a small fraction of the quiet Sun connects into the solar
wind. The quiet 
Sun spectra are also utilised to determine the coronal density and
temperature, leading to average values of $2.6^{+0.5}_{-0.4}\times
10^8$~cm$^{-3}$ and 
$\log\,(T/{\rm K})=5.95\pm 0.02$. No significant trend with the rise
in solar activity during 1996--98 is found for any of the derived
quantities, implying that quiet Sun regions show little dependence on
the solar cycle.
}

\begin{document}

\maketitle

\section{Introduction}\label{sect.intro}

Abundance anomalies in the solar atmosphere and solar wind have been
reported for over 40 years \citep[see, e.g., review of][]{feldman00},
with the general result that elements with a low first ionization
potential (FIP) are enhanced over those with a high FIP. The
separation between low and high FIP is around 10~eV and thus, for
example, Fe, Mg and Si are considered low-FIP elements and Ne and Ar
are considered high-FIP elements. This abundance anomaly pattern is
generally referred to as the ``FIP Effect''. 
The FIP effect is important as it provides a constraint that models
for the solar corona and solar wind must satisfy.

At solar minimum the slow solar wind originates from the quiet Sun and
thus coronal and transition region plasma in the quiet Sun may be
expected to show the FIP enhancement of a factor 4--5 typically found
in the slow solar wind \citep{meyer85}. Measurements of relative
abundances are, 
however, rare for the quiet Sun partly due to limitations of solar
instrumentation. The S082A spectroheliograph on \emph{Skylab} had a slitless
design that resulted in spectral and spatial dimensions being mixed
and only small, isolated features could be studied for abundance work,
ruling out quiet Sun observations on the solar disk. \citet{feld93},
however, were able to use the limb-brightening of transition region
lines at the solar limb to isolate emission from \ion{Ne}{vi} and
\ion{Mg}{vi} ions, yielding a Mg/Ne abundance enhancement over the
photosphere of a factor 1.5--2.0.

\citet{spadaro96} used the average supergranule cell centre and
network region spectra of \citet{vern78} obtained with the S-055
instrument on board \emph{Skylab} to investigate element abundances in
the transition region and corona. They found no enhancement over
photospheric values for the elements considered here (magnesium and
neon), although there was only limited temperature overlap between the
ions they considered.

A key advantage of the NIS of CDS is that it obtains simultaneous
spectral and spatial information for two large wavelength bands,
allowing many ions from different species to be compared -- an
important requirement for abundance studies.
There have been several analyses of CDS data that have yielded abundance
measurements in quiet Sun regions. \citet{landi98} performed a
differential emission measure (DEM) analysis on an active region and a
nearby area of quiet Sun, yielding a Mg/Ne enhancement over the
photospheric value of 2.0 for the latter. \citet{macpher99} studied
the intensities of helium lines in the quiet Sun and a by-product of
their analysis was that the low-FIP enhancement over high-FIP elements
was at most a factor 1.4, with a small variation between
supergranule cell centres and network regions. 
\citet{young98} studied quiet Sun spectra averaged over many CDS
rasters from 1996 and determined a Mg/Ne ratio enhanced over the
photospheric value by a factor 2, with little difference between cell
centres and network. It is to be noted that the present work differs
from \citet{young98} through the analysis method used, and the
separate study of individual data-sets over time.

All of the results from these works refer to the transition
region. Results for the corona are not common in the
literature. \citet{laming95} re-analysed an EUV rocket spectrum of
\citet{mal73} to determine a FIP enhancement of 3--4. Above the solar
limb, \citet{feld98} have measured a low-FIP enhancement of a factor 4
using the SUMER instrument on board SOHO at a height of
1.03~$R_{\odot}$. Another effect seen further off-limb above the quiet
Sun is that of gravitational settling of heavier
elements. \citet{raymond97} used SOHO/UVCS spectra obtained at
1.5~$R_{\odot}$ to demonstrate that oxygen is underabundant relative
to hydrogen by an order of magnitude; for ions of comparable mass,
however, the FIP effect is still seen with a value around 4.

The present work uses spectra from CDS to determine Mg/Ne relative
abundance ratios in the quiet Sun for a number of data-sets over a
period of 28 months 
from the most recent solar activity minimum (1996) through the rise in
activity during 1997 and 1998. The aim is to
provide definitive measurements for supergranule cell centres and
network regions and to search for any long-term time variability.

%% Previous work on the FIP effect in the quiet Sun has focussed on
%% either the transition region ($T< 8\times 10^5$~K) or the corona ($T>
%% 8\times 10^5$~K). The present work presents results for the transition
%% region. \citet{feld93} study limb-brightening rings in emission lines
%% of \ion{Ne}{vi} and \ion{Mg}{vi} to derive Mg/Ne enhancements of 1.5--2.0.

%% Few results on the FIP effect have been reported for quiet Sun
%% regions, partly due to limitations of solar instrumentation. Abundance
%% estimates require emission lines from ions of two different elements
%% that are formed at the same temperature, and often such lines are
%% widely dispersed in wavelength. Thus the SUMER instrument on SOHO and
%% the Harvard S055 instrument on SMM

\section{Observations}

The Coronal Diagnostic Spectrometer (CDS) is one of 12 scientific
instruments on board the Solar and Heliospheric Observatory (SOHO) and
obtains extreme ultraviolet (EUV) 
spectra over the range 150--800\,\AA\ \citep{harrison95}. It consists of two separate
spectrometers (the Grazing Incidence, GIS, and Normal Incidence, NIS)
fed by the same telescope. Only data 
from the NIS are presented here. 
NIS simultaneously obtains spectral and one-dimensional spatial
information in single exposures, with images of a region built up
through rastering. Two wavebands are observed simultaneously:
308--381\,\AA\ and 513--633\,\AA. A rich selection of emission lines
from different ion species are found in these bands, covering
temperatures from $2\times 10^4$ to $6\times 10^6$~K. 
%Of particular
%relevance to the present study are emission lines of \ion{Mg}{v--viii}
%and \ion{Ne}{iv--vii}.

The data used here are from the NISAT\_S CDS observing sequence that obtains
complete NIS spectra over a spatial area
20\asec$\times$240\asec. NISAT\_S has 
been run once per week on quiet Sun regions for the duration of the
SOHO mission as part of 
the CDS synoptic programme. Only studies run up to the
(temporary) loss of SOHO 
in 1998 June are considered here since CDS line profiles were
modified following the loss 
making the weak magnesium lines difficult to measure.

\section{Data preparation}

A list of all runs of the NISAT\_S study between the start of science
observations (1996 February) and 1998 June 23 was extracted. Through
fields in the headers 
of the data files, only those studies positioned within
300\asec\ of Sun centre, and not located on an active region were
selected. The closest EIT 195\footnote{EIT: Extreme Ultraviolet
  Imaging Telescope, one of the instruments on SOHO. The 195 filter of
  EIT 
  selects only radiation close to 195~\AA\ and images are dominated by
\ion{Fe}{xii} \lam195.1 emission formed at around 1.5 million K.} image in time was extracted from the
SOHO databases, and the position of the NISAT raster was overplotted
allowing the location relative to active regions to be determined.
Only rasters lying in clear quiet Sun patches, and not overlying
strong bright points were selected.
In all, 24 NISAT rasters were chosen. Each were processed with the latest
version of the CDS calibration software. The NIS spectra appear tilted
on the detector and this was corrected to ensure that each emission
line was spatially aligned. Cosmic rays were flagged with the
CDS\_NEW\_SPIKE IDL routine and were not included in the analysis. The
data were calibrated in both photon event 
units and intensity (erg cm$^{-2}$ s$^{-1}$ sr$^{-1}$) units. The
errors on the intensity are derived from the photon event values
following \citet{thompson00}. The intensities of some of the stronger
lines in the NIS2 wavelength band are affected by the narrow slit
burn-in correction \citep[see Sect.~7.7.1 of][]{lang02} that results
in line widths (and thus line intensities) falling by up to 10~\%\
over the period 1996--98. The 
IDL routine 
GET\_WIDTH\_CORR in the CDS analysis software corrects for this effect.
For the
neon lines used in the abundance analysis here, the effects are less
than 2~\%\ over the 
time period considered, and are not accounted for. For the stronger
\ion{O}{v} \lam629.7 and \ion{Mg}{x} \lam624.9
lines considered in Sect.~\ref{sect.other}, however, the intensities
can be affected by up to 10~\%\ and have been corrected for.

\begin{figure}[h]
\includegraphics[scale=0.55]{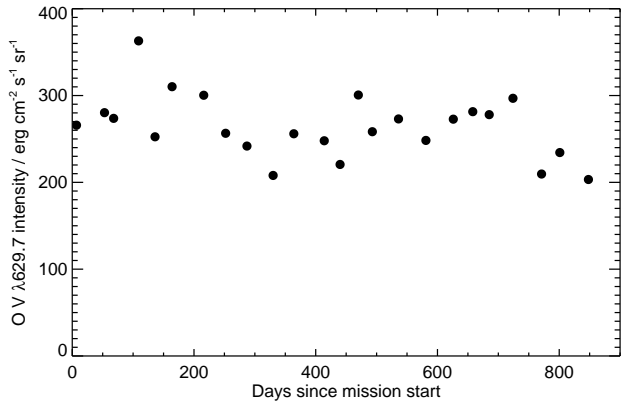}
\caption{The \ion{O}{v} \lam629.7 intensity at which the division between
network and cell centre regions occurs for each of the 24 data-sets.}
\label{fig.o5-half}
\end{figure}

After the data had been calibrated, the spatial separation into supergranule
network and cell centre regions was performed using the \ion{O}{v}
\lam629.7 line. This line is formed at $\log\,(T/{\rm K})=5.4$  and is the
strongest transition region line seen by CDS, giving sufficient signal
at each spatial pixel to allow a direct separation into network or
cell centre regions. 
Based on the work of \citet{gallagher98}, 50 \% of the quiet Sun is found
in network regions and 50 \% in cell centres at the
temperature of formation of \ion{O}{v}. The spatial pixels in
the NISAT rasters were
thus divided according to intensity into network and cell centre
regions. The intensity value at which the division between network and cell
centres occurs for each of the data-sets is shown in
Fig.~\ref{fig.o5-half}. For this and other plots in this paper, time
is represented as the number of days since the start of science
operations for CDS (1996 February 25).
All spatial pixels belonging to the two regions were then
summed, and complete NIS spectra extracted.

%% From the spectra, the emission lines of the magnesium and neon ions
%% were identified and fit with 

A Gaussian fitting routine of the author
based on the MPFIT curve-fitting procedures of C.B. Markwardt\footnote{http://astrog.physics.wisc.edu/$\sim$craigm/idl/idl.html.} was used
to fit lines in the spectrum. Line intensity errors derived from the
fitting procedure
were added in quadrature to flux calibration 
uncertainties, taken as 15~\%.

\section{Mg and Ne ions}

Within the two NIS wavebands, CDS observes emission lines from the
consecutive ion sequences \ion{Mg}{v--viii} and \ion{Ne}{iv--vii}, for
which there is significant temperature overlap
(Fig.~\ref{fig.mg-ne}). Thus the Mg and Ne ions will be found in the
same solar structures and, once the atomic physics parameters are
corrected for, the ratios of the emission lines from the two elements
allow a measure of the relative abundance of the two elements. By
simulaneously comparing four ions from each element, errors due to
atomic data uncertainties in any single ion are avoided.

The transitions used in the present work are listed in
Table~\ref{tbl.lines}. Whereas most of the lines are unblended and
simply fit with a single Gaussian, the \ion{Mg}{v} and \ion{Mg}{vii}
lines 
are more difficult. The former lies close to the stronger
\ion{Fe}{xi} \lam352.7 
and \ion{Fe}{xii} \lam352.1 lines and all three were fitted together
in a 3-Gaussian fit. In cases where the \ion{Mg}{v} line was
particularly weak the widths of all three lines were forced to be the
same. The \ion{Mg}{vii} \lam367.7 line is partially blended with the stronger
\ion{Mg}{ix} \lam368.1 line and the lines were simultaneously fit with
two Gaussians forced to have the same width. Note that all lines
observed with NIS are dominated by instrumental broadening and thus
lines close in wavelength typically have the same width.

\begin{table}[h]
\caption{Emission lines used in the present analysis. Transitions
  within 0.4~\AA\ are blended in the CDS spectra. Wavelengths are from
  v4.2 of the CHIANTI database.}
\begin{flushleft}
\begin{tabular}{lll}
\hline
\hline
\noalign{\smallskip}
Ion &Transition &$\lambda$/\AA  \\
\hline
\noalign{\smallskip}
\ion{Mg}{v}&2s$^2$ 2p$^4$ $^3$P$_{2}$ -- 2s 2p$^5$ $^3$P$_{2}$&353.09\\
&2s$^2$ 2p$^4$ $^3$P$_{1}$ -- 2s 2p$^5$ $^3$P$_{1}$&353.30\\
\noalign{\smallskip}
\ion{Mg}{vi}&2s$^2$ 2p$^3$ $^2$D$_{5/2}$ -- 2s 2p$^4$ $^2$D$_{3/2}$&349.11\\
&2s$^2$ 2p$^3$ $^2$D$_{3/2}$ -- 2s 2p$^4$ $^2$D$_{3/2}$&349.12\\
&2s$^2$ 2p$^3$ $^2$D$_{5/2}$ -- 2s 2p$^4$ $^2$D$_{5/2}$&349.16\\
&2s$^2$ 2p$^3$ $^2$D$_{3/2}$ -- 2s 2p$^4$ $^2$D$_{5/2}$&349.18\\
\noalign{\smallskip}
\ion{Mg}{vii}&2s$^2$ 2p$^2$ $^3$P$_{2}$ -- 2s 2p$^3$ $^3$P$_{2}$&367.66\\
&2s$^2$ 2p$^2$ $^3$P$_{2}$ -- 2s 2p$^3$ $^3$P$_{1}$&367.67\\
\noalign{\smallskip}
\ion{Mg}{viii}&2s$^2$ 2p $^2$P$_{3/2}$ -- 2s 2p$^2$ $^2$P$_{3/2}$&315.02\\
\noalign{\smallskip}
\ion{Ne}{iv}&2s$^2$ 2p$^3$ $^4$S$_{3/2}$ -- 2s 2p$^4$ $^4$P$_{5/2}$&543.89\\
\noalign{\smallskip}
\ion{Ne}{v}&2s$^2$ 2p$^2$ $^3$P$_{2}$ -- 2s 2p$^3$ $^3$D$_{1}$&572.03\\
\noalign{\smallskip}
\ion{Ne}{vi}&2s$^2$ 2p $^2$P$_{3/2}$ -- 2s 2p$^2$ $^2$D$_{3/2}$&562.70\\
&2s$^2$ 2p $^2$P$_{3/2}$ -- 2s 2p$^2$ $^2$D$_{5/2}$&562.80\\
\noalign{\smallskip}
\ion{Ne}{vii}&2s 2p $^3$P$_{1}$ -- 2p$^2$ $^3$P$_{1}$&561.38\\
&2s 2p $^3$P$_{2}$ -- 2p$^2$ $^3$P$_{2}$&561.73\\
\noalign{\smallskip}
\ion{Si}{ix}&2s$^2$ 2p$^2$ $^3$P$_{1}$ -- 2s 2p$^3$ $^3$D$_{1}$&344.95\\
&2s$^2$ 2p$^2$ $^3$P$_{1}$ -- 2s 2p$^3$ $^3$D$_{2}$&345.12\\
&2s$^2$ 2p$^2$ $^3$P$_{2}$ -- 2s 2p$^3$ $^3$D$_{1}$&349.62\\
&2s$^2$ 2p$^2$ $^3$P$_{2}$ -- 2s 2p$^3$ $^3$D$_{2}$&349.79\\
&2s$^2$ 2p$^2$ $^3$P$_{2}$ -- 2s 2p$^3$ $^3$D$_{3}$&349.86\\
\noalign{\smallskip}
\ion{Mg}{ix}&2s$^2$ $^1$S$_{0}$ -- 2s 2p $^1$P$_{1}$&368.07\\
\noalign{\smallskip}
\ion{Mg}{x}& 2s $^2$S$_{1/2}$ -- 2p $^2$P$_{1/2}$&624.94\\
\noalign{\smallskip}
\ion{Si}{xii}& 2s $^2$S$_{1/2}$ -- 2p $^2$P$_{1/2}$&520.67\\
\hline
\end{tabular}
\end{flushleft}
\label{tbl.lines}
\end{table}

Table~\ref{tbl.lines} also lists other emission lines used in this work.

\section{Mg/Ne abundance}

\begin{figure}[b]
\includegraphics[scale=0.55]{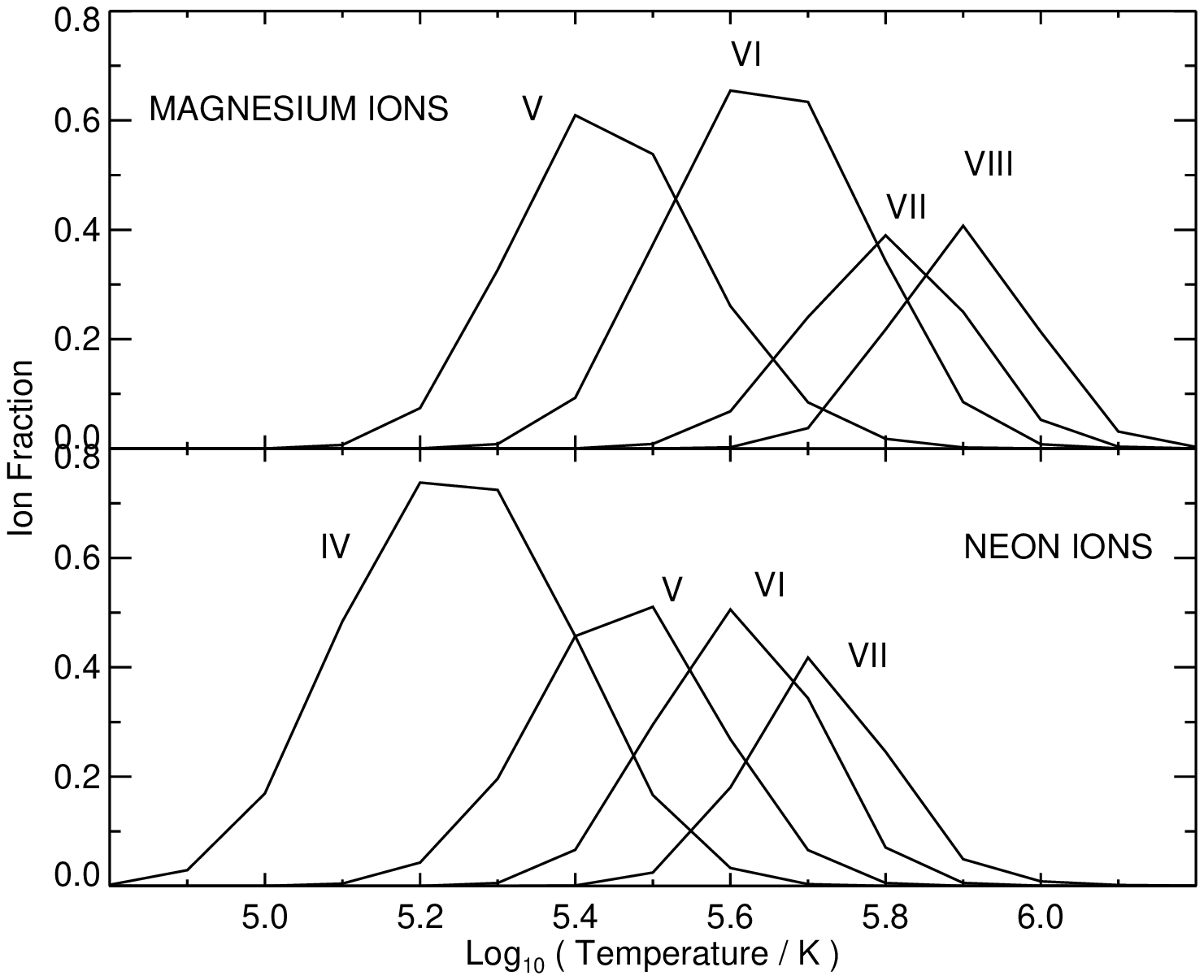}
\caption{Ionization fraction curves from \citet{mazz98} for the
  magnesium 
  and neon ions.}
\label{fig.mg-ne}
\end{figure}

%% The \ion{Mg}{v--viii} and \ion{Ne}{iv-vii} ions have significant
%% temperature overlap over the range $5.2\le\log\,T\le 5.8$
%% (Fig.~\ref{fig.mg-ne}), ensuring 
%% that they are formed in the same plasma. By comparing theoretical
%% intensities with the observed values the Mg/Ne relative abundance can
%% be found.
%% A single emission feature for each ion was considered: \ion{Mg}{v}
%% \lam353.1 (self-blend of two lines), \ion{Mg}{vi} \lam349.2
%% (4 lines), \ion{Mg}{vii} \lam367.7 (2 lines),
%% \ion{Mg}{viii} \lam315.0, \ion{Ne}{iv} \lam543.9, \ion{Ne}{v}
%% \lam572.0, \ion{Ne}{vi} \lam562.8 (2 lines) \ion{Ne}{vii} \lam561.6 (2
%% lines). 

The line fitting procedure resulted in eight intensity values
corresponding to the eight magnesium and neon ions. A minimization
procedure was then applied to derive the Mg/Ne relative abundance
using atomic data from the CHIANTI database and a model for the plasma
distribution with temperature.

The model discretises the plasma into temperature components
distributed at 0.1 dex intervals between $5.0\le \log\,(T/{\rm K})\le
6.1$. This region covers all temperatures at which the Mg and Ne ions
are formed (Fig.~\ref{fig.mg-ne}). The theoretical intensity for an
emission line is given by 
\begin{equation}\label{eq.int}
4\pi I = 0.83\, h\nu\, {\rm Ab}({\rm X}) 
\sum_i n(T_i,N_{\rm e}) \, A \, F(T_i) \, N_{\rm e}(T_i) \, h_i
\end{equation}
where $I$ is the intensity, $h\nu$ the energy for the transition,
Ab(X) the abundance of element X, $n$ the population of the upper
emitting level of the transition (normalised such that the total of
all the level populations
is set to 1), $A$ the radiative
decay rate for the transition, $F$ the ionization fraction, $N_{\rm
  e}$ the electron number density, and $h_i$ the plasma column depth
at temperature $T_i$. The $n$ and $A$ values are derived
from v4.2 of the CHIANTI database \citep{young03} and the $F(T_i)$
values are from \citet{mazz98}. A constant pressure is assumed based
on the density derived from the \ion{Si}{ix} density diagnostic
(Sect.~\ref{sect.other}). The pressure ($=TN_{\rm e}$) is set to
$10^{14.5}$~K~cm$^{-3}$. The effects on the results of assuming
constant density instead are discussed below.

\begin{figure}[h]
\includegraphics[scale=0.55]{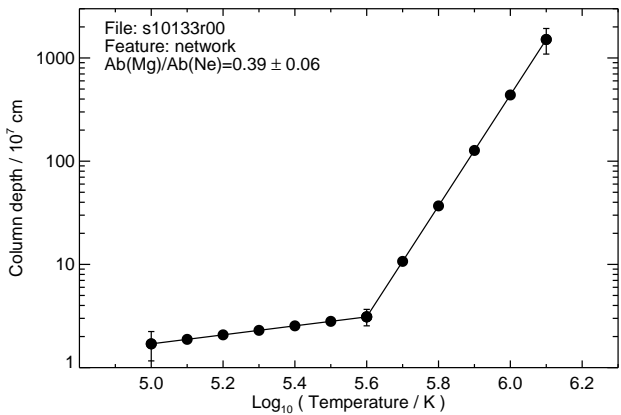}
\caption{Column depths ($h$) derived from the supergranule network
  spectrum of CDS data-set s10133r00 (1998 January 10). The three
  values at $\log\,T$=5.0, 5.6 and 6.1 were derived through the
  minimization procedure, all other values are obtained by linear
  interpolation from these three values.}
\label{fig.h-vals}
\end{figure}

The $h_i$ values are determined through the minimization procedure,
although only three of the values enter as free parameters: the values
at $\log\,T=5.0, 5.6, 6.1$, which we refer to as $h_0$, $h_6$ and
$h_{11}$. The remaining values are determined by linear interpolation in
the $\log\,h$--$\log\,T$ plane. An example of the derived $h$ values
for one data-set is shown in Fig.~\ref{fig.h-vals}.

%% The column depths, $h_i$, are
%% defined over the temperature range 
%% $5.0\le\log\,T\le 6.1$ at 0.1 dex intervals. Only the $h_i$ values at
%% $\log\,T=5.0, 5.6, 6.1$ are free parameters. The remaining values are
%% obtained by linear interpolation in the $\log\,h$--$\log\,T$
%% plane. The electron density is assumed constant with temperature, and
%% we choose a value of $10^9$~cm$^{-3}$. The final results are, however,
%% only weakly sensitive to $N_{\rm e}$. Data from \citet{mazzotta98}
%% were used for the ionization fractions, while the $n$ and $A$ data are
%% from v4.2 of the CHIANTI database \citep{young03}.

\begin{figure}[t]
\includegraphics[scale=0.55]{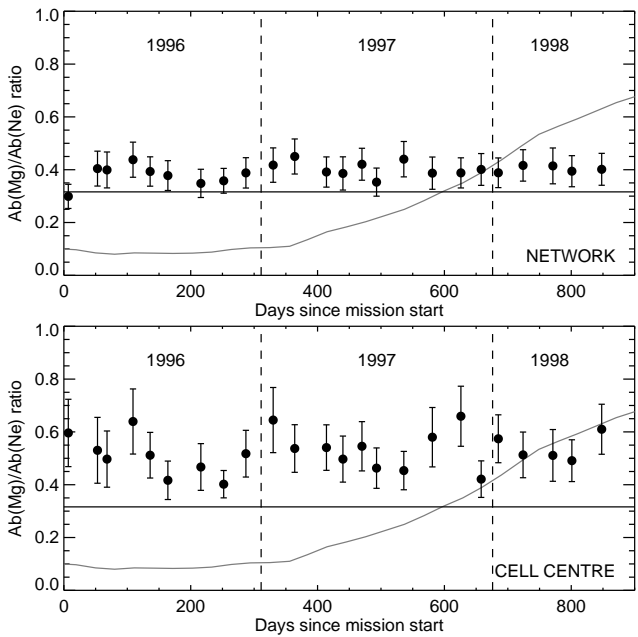}
\caption{The derived Mg/Ne abundance ratios for the network (upper
  panel) and cell
  centre (lower panel) regions as a function of time. The black
  horiontal line denotes the photospheric Mg/Ne abundance ratio. The
  grey line shows the variation in the sunspot number over the period
  of observations.}
\label{fig.ab-ratio}
\end{figure}

The MPFIT procedures mentioned previously were used to minimise the
difference between the theoretical intensities (derived from
Eq.~\ref{eq.int}) and observed
intensities through the four parameters Ab(Mg)/Ab(Ne), $h_0$, $h_6$ and
$h_{11}$. $\chi^2$ values ranged from 0.66 to 3.2 over the different
data-sets.\footnote{The complete set of results from the minimization procedure
is given in Table 2 -- available online.}
The Ab(Mg)/Ab(Ne) values for the network and cell centre regions are
displayed in Fig.~\ref{fig.ab-ratio}. The photospheric Mg/Ne abundance
is taken as 0.316 \citep{grevesse98}. In both types of region there
is no significant trend in time, and the average enhancements over the
photospheric Mg/Ne abundance are $1.25\pm 0.10$ (network) and $1.66\pm
0.23$ (cell centres). The average enhancement of the cell centres over
the network is 
$1.33\pm 0.21$. Also shown in Fig.~\ref{fig.ab-ratio} is the variation
in sunspot number\footnote{Obtained from
  ftp://ftp.ngdc.noaa.gov/STP/SOLAR\_DATA/SUNSPOT\_NUMBERS.} over the
period of the CDS observations. This demonstrates that there is no
trend with solar activity. The abundance results are considered to
apply over the temperature range $5.3\le \log\,(T/{\rm K}) \le 5.8$ as
this the principle range of overlap of the magnesium and neon ions
(Fig.~\ref{fig.mg-ne}).

Assuming a constant density in the minimization procedure produces
only small changes in the results. E.g., for a density of
$10^9$~cm$^{-3}$, the network and cell centre abundance ratios are
$1.25\pm 0.11$ and $1.69\pm 0.25$; while for a density of
$10^{10}$~cm$^{-3}$ they become $1.20\pm 0.13$ and $1.67\pm 0.29$.

In all data-sets the cell centre Mg/Ne ratio is greater than the
network value. The largest enhancement of cell centre over network is
1.99 while the smallest enhancement is 1.03.

The average values for the column depths $h_0$, $h_6$ and $h_{11}$
are: 64.1, 165 and 72,800~km (cell centres); and 157, 350 and
104,000~km (network). Note that these values assume that neon has its
photospheric abundance \citep[$\log\,{\rm Ab}({\rm Ne})/{\rm Ab}({\rm
  H})=-3.92$,][]{grevesse98} in the transition region.

%% \section{Oxygen abundance}

%% Using the column depths derived in the previous section, it is
%% possible to compute theoretical intensities for other emission lines
%% formed in the same temperature range as the Mg and Ne ions. \ion{O}{v}
%% is found at slightly hotter temperatures than \ion{Ne}{iv}

\section{Other quiet Sun properties}\label{sect.other}

Complete NIS spectra were extracted in order to study the Mg/Ne
abundance ratio and other lines in these spectra can be used to
determine additional properties of the quiet Sun. Presented below are
coronal density and temperature measurements, and line intensities as
a function of time.

\begin{figure}[h]
\includegraphics[scale=0.55]{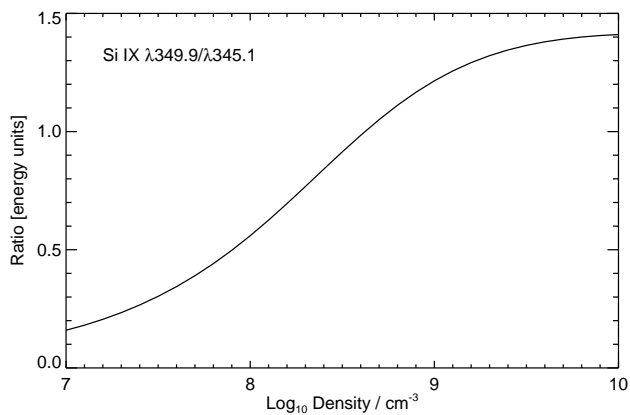}
\caption{Variation of the \ion{Si}{ix} \lam349.9/\lam345.1 ratio as a
  function of density. Calculated with v4.2 of the CHIANTI database at
  a temperature of $\log\,(T/{\rm K})=6.1$.}
\label{fig.si9}
\end{figure}

\begin{figure}[h]
\includegraphics[scale=0.55]{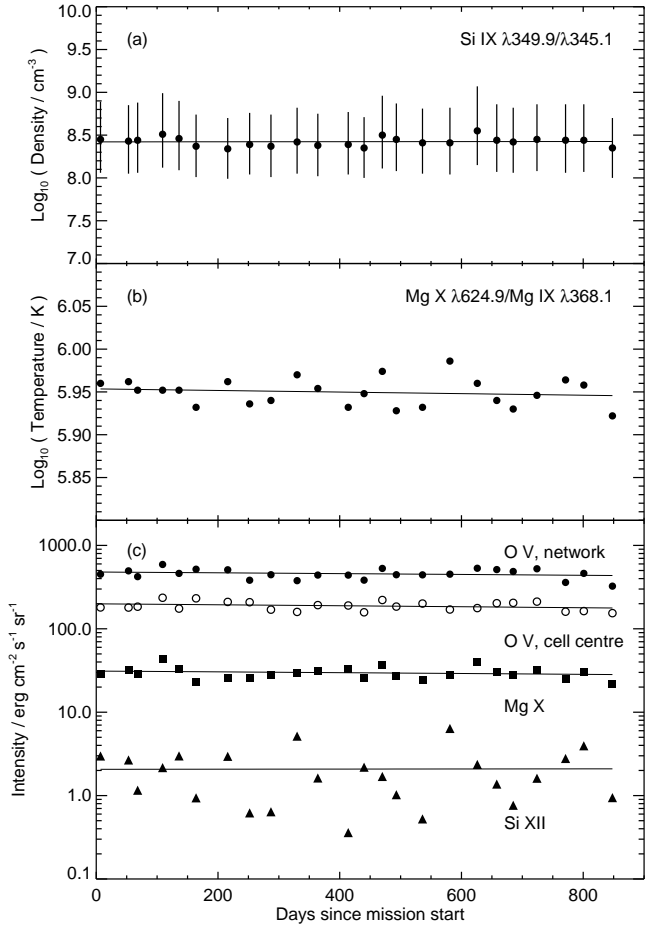}
\caption{Variation of spectral properties over the 850 day period
  considered. (a) 
  electron density derived from \ion{Si}{ix} \lam349.9/\lam345.1
  diagnostic; (b) coronal isothermal temperature derived from
  \ion{Mg}{x} \lam624.9/\lam368.1 ratio; (c) absolute line intensities
  for three species. In each plot linear fits to the data are shown.}
\label{fig.multi-plot}
\end{figure}

Two emission lines of \ion{Si}{ix} are found at 345.1 and 349.9~\AA\
in the NIS spectra and they are blends of 2 and 3 \ion{Si}{ix}
transitions (Table~\ref{tbl.lines}), respectively. The
\lam349.9/\lam345.1 ratio is density sensitive and the range of
sensitivity is particular useful
for coronal hole and quiet Sun regions (Fig.~\ref{fig.si9}). 
\ion{Si}{ix} is formed at temperatures of $\approx$~1 million
K, and so the separation of cell and network regions has limited
physical significance at coronal temperatures. The intensities are
thus averaged to yield quiet Sun intensities, and the
\lam349.9/\lam345.1 line ratio is converted to an electron density
through software in the CHIANTI database. Fig.~\ref{fig.multi-plot}(a)
shows the density values for each of the data-sets. The average value
is $2.6^{+0.5}_{-0.4}\times 10^8$~cm$^{-3}$.

\citet{landi98} found quiet Sun densities of $5.0^{+6.2}_{-2.8}\times
10^8$ 
and $7.9^{+11.1}_{-4.7}\times 10^8$~cm$^{-3}$ using \ion{Si}{ix} for
two quiet Sun regions. We note, however, that the high temperature
\ion{Si}{xii} 
\lam520.7 intensity for these regions is 145 and 254~\ecss,
respectively, 
significantly higher than the values found in the present work (see
below and Fig.~\ref{fig.multi-plot}), indicating that there is some
contamination from the nearby active region.

The ratio of the \ion{Mg}{x} \lam624.9 line to the \ion{Mg}{ix}
\lam368.1 line can be used to yield a temperature estimate for the
quiet Sun if the plasma is assumed to be isothermal. Although the
solar atmosphere is clearly multithermal, the \ion{Mg}{x}/\ion{Mg}{ix} ratio
is useful since the peak of the emission measure distribution in the
quiet Sun is around $\log\,T=5.9$--6.0 \citep{delzanna99}, close to
the temperatures of 
maximum formation of the two ions. Fig.~\ref{fig.multi-plot}(b) shows
the temperatures derived from the line ratio. The values are all very
similar and there is no trend with time. The average value is
$\log\,(T/{\rm K}) =5.95\pm 0.02$.

Fig.~\ref{fig.multi-plot}(c) shows the variation of three emission
lines with time: \ion{O}{v} \lam629.7, \ion{Mg}{x} \lam624.9 and
\ion{Si}{xii} \lam520.7. The latter two are coronal lines and so are
averaged over the network and cell centre regions.  The average
intensities in units 
erg~cm$^{-2}$~s$^{-1}$~sr$^{-1}$ are $458\pm
63$, $189\pm 24$ (\ion{O}{v} network and cell centre), $29.7\pm 5.1$
(\ion{Mg}{x}) and $2.1\pm 1.5$ (\ion{Si}{xii}). The very low intensity of
\ion{Si}{xii} \lam520.7 (one of the stronger lines in active region
spectra) means there is no significant plasma emission above  $\log\,T=6.1$.

%% None of the parameters considered (abundance, density, temperature and
%% line intensities) show a significant trend with the rise
%% in solar activity over the period considered, suggesting that quiet
%% Sun properties remain constant with time. Only the fraction of true
%% quiet Sun regions on the solar surface dminishes with activity.

%\section{Discussion}

\section{Conclusions}

The Mg/Ne abundance ratio in the transition region of the Sun at
temperatures $5.3\le \log\,T\le 5.8$ is enhanced over the value in the
photosphere by less than a factor 2. In supergranule cell centre
regions the enhancement factor is $1.66\pm 0.23$, while for network
regions it is $1.25\pm 0.10$. There is no trend with solar cycle over
the 28 month period (1996 Feb to 1998 Jun) for which data was
analysed.

Simultaneous measurements of the electron density using the density
sensitive \ion{Si}{ix} \lam349.9/\lam345.1 line ratio yield an average
value of $2.6^{+0.5}_{-0.4}\times 10^8$~cm$^{-3}$ for the quiet
Sun. The \ion{Mg}{x} \lam624.9/\ion{Mg}{ix} \lam368.1 ratio yields an
average quiet Sun temperature $\log\,(T/{\rm K}) =5.95\pm 0.02$. 

The Mg/Ne ratios found in this work differ significantly from those
found from in situ measurements in the solar wind, yet are consistent
with the results of the previous works listed in Sect.~\ref{sect.intro}
for the transition region. 
%% The enhancement
%% factor of the measured Mg/Ne ratio over the photospheric value of
%% 0.316 is found to be, on average, $1.25\pm 0.10$  for quiet Sun network regions
%% and $1.66\pm 0.23$ for supergranule cell centres. 
\citet{vonsteiger00} used in situ measurements from the SWICS
instrument on \emph{Ulysses} to determine fast and slow solar wind
element abundances for a number of different elements. Their Mg/Ne
abundance ratio derived for the period 1997 July to 1998 April (which
overlaps with the present work) is a factor $4.4\pm 2.0$ greater than
the photospheric value and much higher than the present values.
%% found a Mg/Ne enhancement factor of $4.4\pm 2.0$
%% in the slow solar wind during the period using in situ measurements
%% from the SWICS 
%% instrument on \emph{Ulysses} 1997 July to 1998 April
%% These compare with the
%% enhancement of $4.4\pm 2.0$  found by \citet{vonsteiger00} 
%% using in situ measurements from the SWICS instrument on \emph{Ulysses}
%% during the period 1997 July to 1998 April. 
Given that, during this
quiet period of the Sun's activity cycle, the slow solar wind must be
accelerated from quiet Sun regions, then a conclusion must be that only
a fraction of the quiet Sun can connect out into the solar
wind. Closed field regions which form the majority of the bright
network appear to show photospheric abundances.

Supergranule cell centres contain less closed field and there is a clear
enhancement of the Mg/Ne ratio over network regions as derived from
the CDS data. This can be interpreted as a greater contribution of
open field line plasma to the total emission from cell
centres. The cell centre abundance values are, however, still
considerably below the slow solar wind Mg/Ne ratio. High spatial
resolution, high sensitivity observations of superganule cell centres
are required to search for small open-field regions that would be
expected to show the high Mg/Ne enhancements found in the solar wind.

None of the quantities (Mg/Ne abundance, density, temperature)
measured in the present work show a significant trend with time as
solar activity increases from its minimum in 1996. In particular, the
hottest line in this work (\ion{Si}{xii} \lam520.7, formed at
$\log\,[T/{\rm K}]=6.3$) shows an average constant trend with
time. This implies that true quiet Sun regions are not affected by the
solar cycle. 
%% The global changes in the Sun's conditions are purely
%% driven by 
%% the increasing number of active regions during the cycle -- the
%% ``quiet Sun'' is physically a ubiquitous.
This is consistent with the relatively small changes in quiet Sun
magnetic flux during the solar cycle \citep{hagenaar03}, which may be
indicative of a separate, solar cycle independent, dynamo mechanism
that generates the ephemeral 
regions found in the quiet Sun.

CDS has allowed a detailed study of the FIP effect in the quiet Sun at
transition region temperatures due to the simultaneous observation of
several ions of magnesium and neon. A similar study for coronal plasma
in the quiet sun is not possible with CDS as there are no useful lines
of sulphur and argon (coronal lines of neon are found at X-ray
wavelengths). However, the \emph{EUV Imaging Spectrometer} to 
be flown on Solar-B \citep{culhane00} will observe consecutive series
of ions from iron and sulphur over 1 to 3 million K. An important
project for EIS will be to measure the FIP enhancement at these
temperatures in the quiet sun and determine whether it is closer to
the solar wind value or the transition value measured here.

\begin{acknowledgements}
The author thanks J.~Schmelz, A.~Fludra and J.~Lang for useful comments.
\end{acknowledgements}

%% \section{Conclusions}

%% Quiet Sun spectra from CDS over the period 1996 February to 1998 June
%% have been analysed, and Mg/Ne abundance ratios for network and
%% supergranule cell centre regions derived. Spectra close to disk centre
%% have been studied to minimize projection effects. The enhancements over
%% photospheric values are $1.25\pm 0.13$ and $1.72\pm
%% 0.22$, respectively. At solar minimum, we expect the slow solar wind
%% to emanate from quiet Sun regions, yet the FIP enhancement in the
%% solar wind is 4--5, much higher than the values here. This can be
%% understood if the slow solar wind comes from quiet Sun regions that
%% have low transition region emission. If we associate network
%% regions with bright, small scale (closed) loops then these loops must
%% have close-to photospheric abundances. The higher Mg/Ne enhancement in
%% cell centre regions is consistent with the FIP effect occurring in low
%% intensity region.

%% The quiet Sun data have also been analysed to yield an average coronal
%% density of $2.7^{+0.6}_{-0.5}\times 10^8$~cm$^{-3}$ through the
%% \ion{Si}{ix} \lam349.9/\lam345.1 density 
%% diagnostic.

% \appendix

% \section{}
% \label{}

% Bibliographic references with the natbib package:
% Parenthetical: \citep{Bai92} produces (Bailyn 1992).
% Textual: \citet{Bai95} produces Bailyn et al. (1995).
% An affix and part of a reference:
%   \citep[e.g.][Ch. 2]{Bar76}
%   produces (e.g. Barnes et al. 1976, Ch. 2).

\end{document}